\documentclass[epj]{svjour}
\usepackage{graphics}
\begin{document}
\title{Probing the exotic structure of $^{8}$B by its elastic
scattering and breakup reaction on nuclear targets}
\author{V.K.~Lukyanov\inst{1} \and D.N.~Kadrev\inst{2} \and E.V.~Zemlyanaya\inst{1}
\and K.V.~Lukyanov\inst{1} \and A.N.~Antonov\inst{2} \and
M.K.~Gaidarov\inst{2} \and K.~Spasova\inst{2,3}
}                     
%
%
\institute{Joint Institute for Nuclear Research, Dubna 141980,
Russia \and Institute for Nuclear Research and Nuclear Energy,
Bulgarian Academy of Sciences, Sofia 1784, Bulgaria \and
University "Ep.~K. Preslavski", Shumen 9712, Bulgaria}
\date{Received: date / Revised version: date}
%
\abstract{The structure of the exotic $^{8}$B nucleus is studied
by means of elastic scattering, as well as its breakup on nuclear
targets. We present microscopic calculations of the optical
potentials (OPs) and cross sections of elastic scattering of
$^{8}$B on $^{12}$C, $^{58}$Ni, and $^{208}$Pb targets at energies
$20<E<170$ MeV. The density distributions of $^{8}$B obtained
within the variational Monte Carlo (VMC) model and the
three-cluster model (3CM) are used to construct the potentials.
The real part of the hybrid OP is calculated using the folding
model with the direct and exchange terms included, while the
imaginary part is obtained on the base of the high-energy
approximation (HEA) and also taken to be equal to the microscopic
real part of the OP. In this model the only free parameters are
the depths of the real and imaginary parts of OP obtained by
fitting the elastic scattering experimental data. It is found a
dependence of their values on the model density of $^{8}$B. In
addition, cluster model, in which $^{8}$B consists of a $p$-halo
and the $^{7}$Be core, is applied to calculate the breakup cross
sections of $^{8}$B nucleus on $^{9}$Be, $^{12}$C, and $^{197}$Au
targets, as well as momentum distributions of $^{7}$Be fragments,
and a comparison with the existing experimental data is made.
} 
\authorrunning{V.K.~Lukyanov et al.}
\titlerunning{Probing the exotic structure of $^{8}$B by its elastic
scattering and breakup reaction ...}

\maketitle
\section{Introduction}
\label{s:intro}

The development of the radioactive ion beams has allowed studies
of nuclei far from stability. This technical headway led to the
discovery of halo nuclei on the neutron-rich side of the valley of
stability \cite{Tanihata85a,Tanihata85b}. These weakly bound
nuclei have a strongly clusterized structure
\cite{Hansen95,Tanihata96,Jonson2004,Baur2003}. In a simple model,
they are seen as a core, that contains most of the nucleons, to
which one or two neutrons are loosely bound. Due to this poor
binding, the valence neutrons tunnel far outside the classically
allowed region and form a sort of halo around the core
\cite{Hansen87}.

Although less probable, proton halos are also possible. Knowledge
about the short-lived radioactive nucleus $^{8}$B is valuable both
for astrophysical reasons \cite{Davids1998,Davids2003,Trache2003}
and for clarifying the question of the existence of proton halo.
Many different findings of evidence argue for the latter. First,
the proton separation energy of only 136.4 keV shows up that
$^{8}$B is the most likely candidate for such a proton-halo
nucleus. Second, the interaction cross section at 790 MeV/nucleon
indicates that the root-mean-square (rms) radius of $^{8}$B is
different from more tightly bound Boron isotopes
\cite{Tanihata88,Obuti96}. The relativistic mean-field
calculations of the rms radii performed systematically for light
isotopes ($A<40$) show that $^{8}$B has a large proton matter
radius compared to its neutron matter radius \cite {Wang2001}. The
enhanced reaction cross section extracted by angular distribution
measurements at intermediate energies
\cite{Negoita96,Fukuda99,Warner95} and the large proton-removal
cross section at relativistic and intermediate energies with
targets ranging from carbon to lead nuclei
\cite{Negoita96,Blank97,Smedberg99} support a halo structure of
$^{8}$B. A radius of the matter density of 2.7 fm, i.e., much
larger than the prediction of any self-consistent calculation is
shown to explain the experimental quadrupole moment of $^{8}$B
\cite{Minamisono92,Sumikama2006}.

The narrow momentum distributions of $^{7}$Be fragments in the
breakup of $^{8}$B measured in C, Al, and Pb targets at 1471
MeV/nucleon with full width at half maximum (FWHM) of $81\pm 6$
MeV/c in all targets have been interpreted in terms of a largely
extended proton distribution for $^{8}$B and have implied a radius
of 2.78 fm \cite{Schwab95}. Here we should mention also the
results of the experiments at lower energies for the breakup of
$^{8}$B in the collisions with Be and Au targets at 41 MeV/nucleon
($81\pm 4$ and $62\pm 3$ MeV/c FWHM for Be and Au targets,
respectively) \cite{Kelley96} and for C target at 36 MeV/nucleon
\cite{Jin2015} with FWHM $124\pm 17$ and $92\pm 7$  MeV/c for the
stripping and diffraction components, correspondingly. Indeed,
these experimental results reflect the large spatial extension of
the loosely bound proton in $^{8}$B. The halo nature of $^{8}$B
nucleus through studies of its breakup has been mostly tested with
cluster models presuming simple two-cluster structure that
consists of $^{7}$Be core and valence proton (for instance,
Refs.~\cite{Smedberg99,Schwab95,Esbensen96}, and
Ref.~\cite{AbuIbrahim2003} presenting a Fortran code to calculate
cross sections for the core plus one-nucleon case) or extended
three-body model ($\alpha$+$^{3}$He+$p$) \cite{Grigorenko98}. The
latter model used to interpret the experimental data in
Ref.~\cite{Smedberg99}, as well as quasiparticle RPA calculations
\cite{Schwab95} and Serber model applied to data in
Ref.~\cite{Kelley96}, demonstrate the necessity to have a rms
$^{7}$Be--$p$ distance of 4--4.5 fm in order to reproduce the
small momentum distribution width. Such a relatively large value
of the rms radius for the last proton in $^{8}$B ($4.20\pm 0.22$
fm) compared to the rms radii of the $^{7}$Be core and $^{8}$B
projectile was also obtained by Carstoiu {\it et al.}
\cite{Carstoiu2001} from the analysis of the experimentally
measured asymptotic normalization coefficient for $^{8}$B
$\rightarrow $ $^{7}$Be+$p$ process. However, as it has been
pointed out in Ref.~\cite{Schwab95}, the ground state of $^{8}$B
is more complex than simply a proton around an inert $^{7}$Be core
and nuclear structure calculations with an appropriate treatment
of continuum effects can describe the measured narrow momentum
distribution and cross sections.

The idea of the existence of a proton halo in $^{8}$B was
experimentally verified also in measurements and studies of
differential cross sections of $^{8}$B elastic scattering on
different nuclear targets in the energy range 20-170 MeV
\cite{Barioni2011,Aguilera2009,Yang2013}. To describe the
elastic-scattering angular distributions and total reaction cross
sections conventional optical-model with Woods-Saxon (WS)
potentials or double-folding (DF) OPs calculations were performed.
The effect of breakup on the elastic scattering was investigated
for the weakly bound $^{8}$B nucleus by performing continuum
discretized coupled-channels (CDCC) calculations
\cite{Barioni2011,Aguilera2009,Yang2013,Paes2012,Lubian2009,Nunes1998,Mukeru2015}.
It was found that for the light $^{12}$C target this effect is
negligible for description of the elastic scattering data
\cite{Barioni2011,Paes2012}. Similar conclusion has been drawn in
Ref.~\cite{Yang2013}, in which the CDCC calculations have shown a
small effect of breakup channel couplings on the
$^{8}$B+$^{208}$Pb elastic scattering at an incident energy well
above the Coulomb barrier.

The aims of the present work are as follows. First, we analyze the
data on elastic scattering cross sections of $^{8}$B on $^{12}$C,
$^{58}$Ni, and $^{208}$Pb targets at energies $20<E<170$ MeV
within the microscopic model of the respective OP and compare the
results with the available experimental data. As in our previous
works
\cite{Lukyanov2007,Lukyanov2009,Lukyanov2010,Lukyanov2013,Lukyanov2015},
where processes with neutron-rich He, Li, and Be isotopes were
considered, we use the hybrid model of OP \cite{Lukyanov2004}, in
which the real part (ReOP) is calculated by a folding of a nuclear
density and the effective nucleon-nucleon (NN) potentials
\cite{Khoa2000} (see also \cite{Lukyanov2007a}) and includes
direct and exchange isoscalar and isovector parts. The imaginary
part (ImOP) is obtained in two ways: i) on the base of the
high-energy approximation method developed in
Refs.~\cite{Glauber,Sitenko} and ii) taken to be equal to
microscopically calculated folding real part of the OP. There are
only two fitting parameters in the hybrid model. They are related
to the depths of the ReOP and ImOP. In the present work we use the
density distribution obtained within the variational Monte Carlo
model \cite{Carlson2015,Pieper2015} and also the density obtained
within the framework of the microscopic three-cluster model of
Varga {\it et al.} \cite{Varga95}. The main effort is to minimize
the ambiguities in the fitted OPs by studying differential elastic
cross sections and to compare them with the available experimental
data, namely for reactions $^{8}$B+$^{12}$C at 25.8 MeV
\cite{Barioni2011}, $^{8}$B+$^{58}$Ni at 20.7, 23.4, 25.3, 27.2,
and 29.3 MeV \cite{Aguilera2009}, and $^{8}$B+$^{208}$Pb at 170.3
MeV \cite{Yang2013}. Second, in addition to the analysis of
elastic scattering cross sections, we estimate important
characteristics of the reactions with $^{8}$B, such as the breakup
cross sections and momentum distributions of fragments in breakup
processes on nuclear targets. For a more consistent description of
the possible halo structure of $^{8}$B we calculate the momentum
distributions of $^{7}$Be fragments from the breakup reactions
$^{8}$B+$^{9}$Be, $^{8}$B+$^{12}$C, and $^{8}$B+$^{197}$Au for
which experimental data are available. Such a complex study based
on the microscopic method to obtain the OPs with a minimal number
of free parameters and by testing density distributions of $^{8}$B
which reflect its proton-halo structure (in contrast, e.g., to the
Hartree-Fock density used in Ref.~\cite{Yang2013}) would lead to a
better understanding of the $^{8}$B structure and to a reduction
of the inconsistency of describing the available data. Also, it
would be a test of our microscopic approach to reveal the
proton-halo structure of the loosely bound $^{8}$B projectile,
particularly its density distribution, in the considered elastic
scattering and breakup processes.

The article is organized as follows. The theoretical scheme to
calculate the real and imaginary parts of the OP, as well as the
results for the $^{8}$B+$^{12}$C, $^{8}$B+$^{58}$Ni, and
$^{8}$B+$^{208}$Pb elastic scattering differential cross sections
are presented and discussed in Sec.~\ref{s:theory}.
Section~\ref{s:theorybu} contains the basic expressions to
estimate the $^{8}$B breakup and results for the fragment momentum
distributions of $^{7}$Be in the stripping process of $^{8}$B on
$^{9}$Be, $^{12}$C, and $^{197}$Au. The summary of the work and
conclusions are given in Sec.~\ref{s:conclusions}.

\section{Elastic scattering of $^{8}$B on $^{12}$C, $^{58}$Ni, and
$ ^{208}$Pb}
\label{s:theory}
\subsection{The microscopic optical potential}
\label{s:op}

The microscopic volume OP used in our calculations contains the
real part ($V^{DF}$) including both the direct and exchange terms
and the HEA microscopically calculated imaginary part ($W^{H}$).
It has the form
\begin{equation}
U(r) = N_R V^{DF}(r) + i N_I W^{H}(r).
\label{eq:1}
\end{equation}
The parameters $N_R$ and $N_I$ entering Eq.~(\ref{eq:1})
renormalize the strength of OP and are fitted by comparison with
the experimental cross sections. Details of the constructing of
the OP are given in
Refs.~\cite{Khoa2000,Lukyanov2007a,Satchler79,Avrigeanu2000}. The
real part $V^{DF}$ consists of the direct ($V^{D}$) and exchange
($V^{EX}$) double-folding integrals that include effective $NN$
potentials and density distribution functions of colliding nuclei.
The $V^{D}$ and $V^{EX}$ parts of the ReOP have isoscalar (IS) and
isovector (IV) contributions. The IS ones of both terms are:
\begin{equation}
V^{D}_{IS}(r) = \int d^3 r_p d^3 r_t  {\rho}_p({\bf r}_p) {\rho}_t
({\bf r}_t) v_{NN}^D(s),
\label{eq:2}
\end{equation}
\begin{eqnarray}
V^{EX}_{IS}(r)&=& \int d^3 r_p d^3 r_t   {\rho}_p({\bf r}_p, {\bf
r}_p+ {\bf s})  {\rho}_t({\bf r}_t, {\bf r}_t-{\bf s})
\nonumber \\
& & \times  v_{NN}^{EX}(s)  \exp\left[ \frac{i{\bf K}(r)\cdot
s}{M}\right],
\label{eq:3}
\end{eqnarray}
where ${\bf s}={\bf r}+{\bf r}_t-{\bf r}_p$ is the vector between
two nucleons, one of which belongs to the projectile and another
one to the target nucleus. In Eq.~(\ref{eq:2}) $\rho_p({\bf r}_p)$
and $\rho_t({\bf r}_t)$ are the densities of the projectile and
the target, respectively, while in Eq.~(\ref{eq:3}) $\rho_p({\bf
r}_p, {\bf r}_p+{\bf s})$ and $\rho_t({\bf r}_t, {\bf r}_t-{\bf
s})$ are the density matrices for the projectile and the target
that are usually taken in an approximate form \cite{Negele72}. The
effective $NN$ interactions $v_{NN}^{D}$ and $v_{NN}^{EX}$ have
their IS and IV components in the form of M3Y interaction obtained
within $g$-matrix calculations using the Paris NN potential
\cite{Khoa2000}. The expressions for the energy and density
dependence of the effective $NN$ interaction are given, e.g., in
Ref.~\cite{Lukyanov2015}.

It is important to note that the energy dependence of $V^{EX}$
arises primarily from the contribution of the exponent in the
integrand of Eq.~(\ref{eq:3}). Indeed, there the local
nucleus-nucleus momentum reads
\begin{equation}
K(r)=\left \{\frac{2Mm}{\hbar^2}\left[E-V^{DF}(r)-V_c(r)\right
]\right \}^{1/2}
\label{eq:4}
\end{equation}
with $M=A_pA_t/(A_p+A_t)$, where $A_{p}$, $A_{t}$, $m$ are the
projectile and target atomic numbers and the nucleon mass. As can
be seen, $K(r)$ depends on the folding potential $V^{DF}(r)$ that
has to be calculated itself and, therefore, we have to deal with a
typical non-linear problem.

Concerning the ImOP, we take it in two forms. One of them
corresponds to the full microscopic OP derived in
Refs.~\cite{Lukyanov2004,Shukla2003} within the HEA
\cite{Glauber,Sitenko}:
\begin{eqnarray}
W^H(r) &=& -\frac{1}{2\pi^2}\frac{E}{k}{\bar\sigma}_{N} \nonumber \\
& \times & \int_0^{\infty} j_0(qr){\rho}_p(q){\rho}_t(q){f}_N(q)
q^2dq.
\label{eq:5}
\end{eqnarray}
In Eq.~(\ref{eq:5}) $\rho(q)$ are the corresponding form factors
of the nuclear densities, $f_N(q)$ is the amplitude of the NN
scattering and $\bar\sigma_{N}$ is the averaged over the isospin
of the nucleus total NN scattering cross section that depends on
the energy and accounts for the in-medium effect
\cite{Charagi92,Shukla2001,Xiangzhow98}. The second form of $W$ is
equal to the microscopically calculated folding real part $V^{DF}$
of the OP.

\subsection{Results of calculations of elastic scattering cross sections}
\label{s:calculations}

We calculate the OP [Eq.~(\ref{eq:1})] and the elastic scattering
differential cross sections of $^{8}$B on different targets using
the DWUCK4 code \cite{DWUCK}. All the elastic scattering cross
sections will be shown in the figures as ratios to the Rutherford
cross section.

\begin{figure}
\resizebox{0.45\textwidth}{!}{%
  \includegraphics{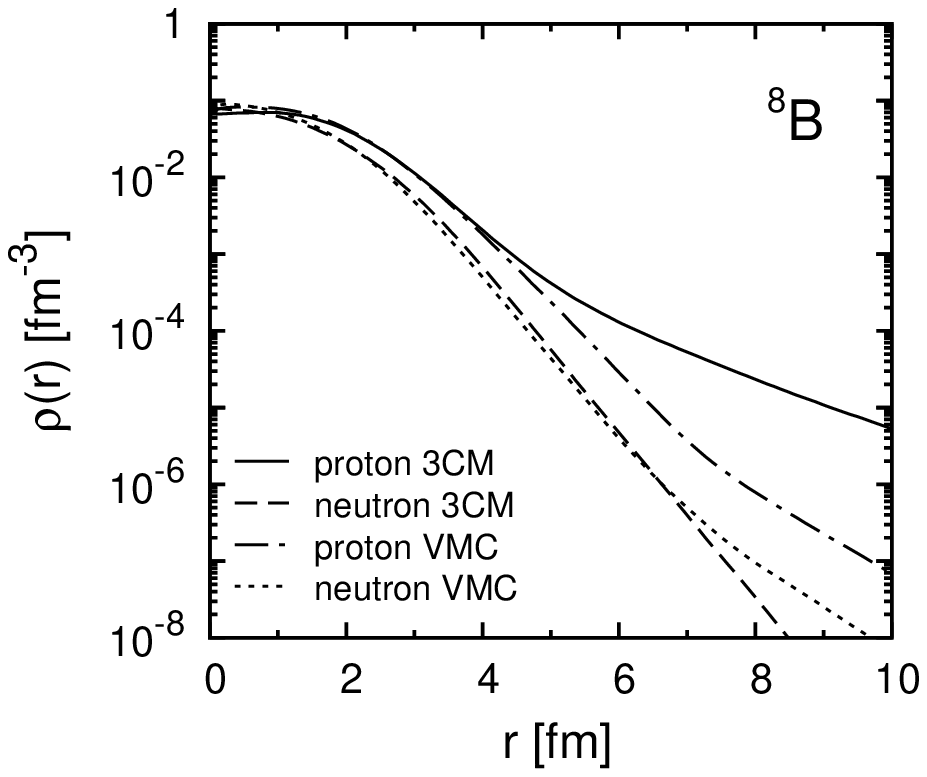}
}
\resizebox{0.45\textwidth}{!}{%
  \includegraphics{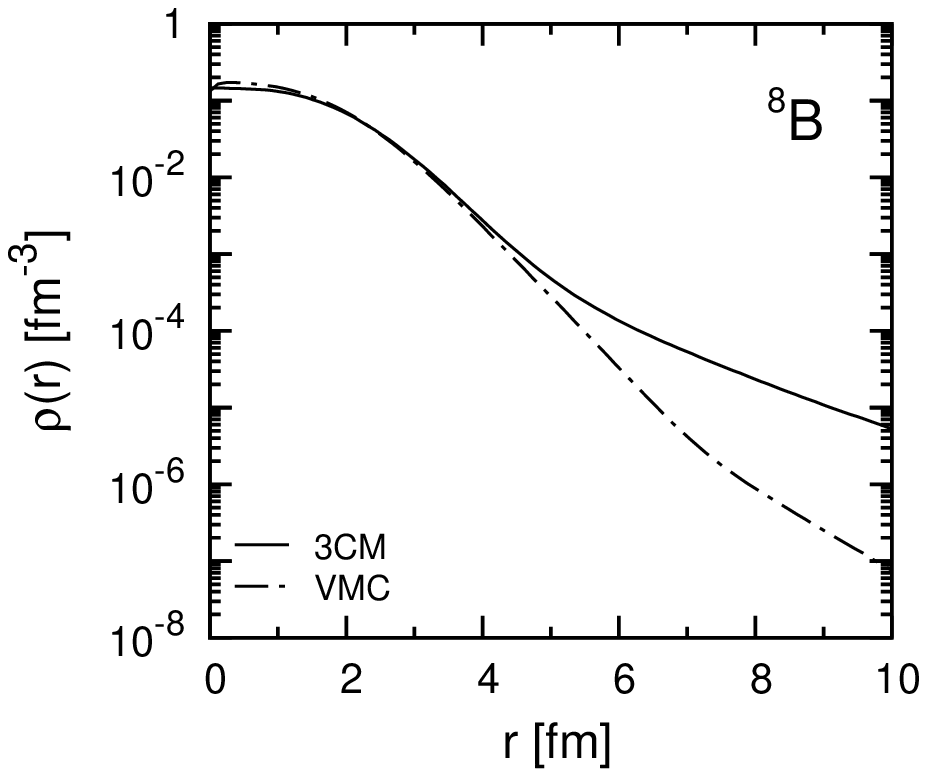}
}
\caption{Point-proton (normalized to $Z$=5), point-neutron
(normalized to $N$=3) (top panel) and the total densities (bottom
panel) of $^{8}$B (normalized to $A$=8) obtained in the VMC
\protect\cite{Carlson2015,Pieper2015} method and in the 3CM
\protect\cite{Varga95}.}
\label{fig1}
\end{figure}

To apply the microscopic OPs to scattering of $^{8}$B on nuclei we
used realistic density distributions of $^{8}$B calculated within
the VMC model \cite{Carlson2015,Pieper2015} and from the 3CM in
Ref.~\cite{Varga95}. In our case, within the VMC method the proton
and neutron densities have been computed with the AV18+UX
Hamiltonian, in which the Argonne v18 two-nucleon and Urbana X
three-nucleon potentials are used \cite{Pieper2015}. Urbana X is
intermediate between the Urbana IX and Illinois-7 models (the
latter was used by us in Ref.~\cite{Lukyanov2015} for the
densities of $^{10}$Be nucleus). As far as the 3CM densities of
Varga {\it et al.} \cite{Varga95} are concerned, the $^{8}$B
nucleus has been studied in a microscopic $\alpha$+$h$+$p$
three-cluster model ($h$=$^{3}$He) using the stochastic
variational method, where a Minnesota effective two-nucleon
interaction composed from central and spin-orbit parts was used.
It has been shown in \cite{Varga95} that the proton separation
energy of $^{8}$B is reasonably reproduced, but the calculated
point matter radius exceeds the "empirical" one. The VMC and 3CM
densities are given in Fig.~\ref{fig1}. It can be seen that they
have been calculated with enough accuracy up to distances much
larger than the nuclear radius. In both methods the total
densities of $^{8}$B occur quite similar up to $r\sim 4$ fm and a
difference between them is seen in their periphery. Due to the
cluster-structure model of $^{8}$B, where the proton is considered
as a single cluster \cite{Varga95}, the tail part of the
point-proton distribution of $^{8}$B is significantly larger than
that of the neutron one, causing considerable difference in the
corresponding rms radii listed in Table~\ref{tab1}. In it we give
also the "empirical" data, e.g. from
Refs.~\cite{Tanihata88,Fukuda99}, for the effective rms radii of
the point-proton, point-neutron and matter distributions deduced
from the Glauber analysis of the interaction and reaction cross
sections. One can see from Table~\ref{tab1} that the values of the
proton and matter rms radii in the case of VMC density are close
to the "empirical" values for $^{8}$B, while the neutron radius
obtained in the 3CM almost coincides with the experimental value.
In the calculations of the OPs for $^{8}$B+$^{12}$C the density of
$^{12}$C was taken in symmetrized Fermi form with radius and
diffuseness parameters 3.593 fm and 0.493 fm \cite{Burov77},
respectively. For $^{8}$B+$^{58}$Ni and $^{8}$B+$^{208}$Pb elastic
scattering the densities of $^{58}$Ni and $^{208}$Pb were taken in
a form of two-parameter Fermi distributions with radius and
diffuseness parameters 4.08 fm and 0.515 fm \cite{Khoa2000}, 6.654
fm and 0.475 fm \cite{Patterson2003}, respectively.

The real part of the OPs in the considered cases are calculated
using Eqs.~(\ref{eq:1})-(\ref{eq:4}). The imaginary part of the
OPs is obtained in two ways: i) $W=W^{H}$, where $W^{H}$ is from
Eq.~(\ref{eq:5}), or ii) taken to be equal to the real part of the
OP ($W=V^{DF}$). It is seen from our results that the different
behavior of the two densities (VMC and 3CM) used in the OPs
calculations reflects on the shape of the ReOP and ImOP in their
periphery being with a longer tail for 3CM density. At the same
time (e.g., in the case of $^{8}$B+$^{58}$Ni elastic scattering),
the imaginary parts of the OPs calculated in HEA and corresponding
to both $^{8}$B densities are almost one order of magnitude deeper
than the real parts. It is clear that this behavior of the central
part of the HEA ImOP is not realistic. From other side, however,
it is known that the decisive region of the OP at such energies is
the surface one. So, our further calculations of the elastic cross
sections explore namely the role of the surface component of the
OP. Further in our work we show, as an example (see
Fig.~\ref{fig7}), our results that confirm this fact. We note, as
it has been pointed out in Ref.~\cite{Aguilera2009}, that every
acceptable potential has an imaginary part that is extended beyond
the corresponding real part. As a result, an absorption at a large
distance due to the existence of a halo state is suggested.

\begin{table}
\caption{Proton, neutron, and matter rms radii (in fm) of $^{8}$B
obtained within the VMC method \cite{Pieper2015} and 3CM
\cite{Varga95}. The "empirical" effective rms radii are from
Refs.~\cite{Tanihata88,Fukuda99}.}
\label{tab1}
\begin{center}
\begin{tabular}{llll}
\hline\noalign{\smallskip}
&   $r_{p}$  & $r_{n}$  & $r_{m}$ \\
\noalign{\smallskip}\hline\noalign{\smallskip}
VMC                & 2.45    & 2.14    & 2.34  \\
3CM                & 2.73    & 2.24    & 2.56  \\
\cite{Tanihata88}  & 2.45    & 2.27    & 2.38  \\
\cite{Fukuda99}    & 2.53    & 2.31    & 2.45  \\
\noalign{\smallskip}\hline
\end{tabular}
\end{center}
\end{table}

In our work we consider the set of the $N_{i}$ coefficients ($N_R$
and $N_I$, see Eq.~(\ref{eq:1}) for the OP) as parameters to be
found out from the fit to the experimental data for the cross
sections using the $\chi^2$-procedure. We should mention (as it
had been emphasized in our previous works
\cite{Lukyanov2007,Lukyanov2009,Lukyanov2010,Lukyanov2013,Lukyanov2015})
that we do not aim to find a complete agreement with the data. The
fitted $N$s related to the depths of the ReOP and ImOP can be
considered as a measure of deviations of our microscopic OPs from
the case when the values of $N$s are equal to unity.

\begin{figure}
\resizebox{0.45\textwidth}{!}{%
  \includegraphics{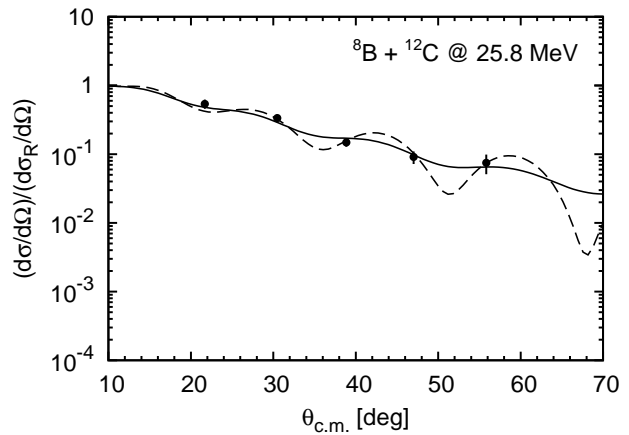}
} \caption{$^{8}$B+$^{12}$C elastic scattering cross sections at
$E=25.8$ MeV. Solid line: calculations with 3CM density of
$^{8}$B; dashed line: calculations with VMC density of $^{8}$B.
Experimental data are taken from Ref.~\protect\cite{Barioni2011}.}
\label{fig2}
\end{figure}

The calculated within the hybrid model elastic scattering cross
sections of $^{8}$B+$^{12}$C at energy $E=25.8$ MeV in the
laboratory frame are given in Fig.~\ref{fig2} and compared with
the experimental data \cite{Barioni2011}. It can be seen that in
both cases of calculations with VMC or 3CM densities the results
are in good agreement with the available data. The differential
cross section obtained with VMC density demonstrates more
developed diffractional picture. It would be desirable to measure
the elastic scattering in the angular range beyond 55$^\circ$,
where the differences between the theoretical results start, in
order to determine the advantage of using VMC or 3CM microscopic
densities of $^{8}$B. Complementary measurements at smaller steps
of scattering angle would also allow one to observe some possible
oscillations of the cross section. Our results reproduce the
experimental data better than the analysis of the data
\cite{Barioni2011} with optical-model calculations using S\~{a}o
Paulo potential with energy dependence and nonlocality correction
\cite{Chamon2002}. The imaginary part of the latter potential has
the same form factor as the real part that has not been
renormalized ($N_{R}=1$), but with a normalization of
$N_{I}=0.78$. The corresponding values of $N_{R}$ and $N_{I}$
parameters obtained in our work for 3CM and VMC densities, as well
as the total reaction cross sections $\sigma_{R}$ (in mb) are
listed in Table~\ref{tab2}.

\begin{table}
\caption{The renormalization parameters $N_{R}$, $N_{I}$, and the
total reaction cross sections $\sigma_{R}$ (in mb) for results of
the $^{8}$B+$^{12}$C and $^{8}$B+$^{208}$Pb elastic scattering
processes at incident energy $E$ (in MeV) considered and shown in
Figs.~\ref{fig2} and \ref{fig3} using the 3CM and VMC model
densities of $^{8}$B, respectively.}
\label{tab2}
\begin{center}
\begin{tabular}{lllllll}
\hline\noalign{\smallskip}
Process & Model & $E$ & $N_R$ & $N_I$ & $\sigma_R$\\
\noalign{\smallskip}\hline\noalign{\smallskip}
$^{8}$B+$^{12}$C   & 3CM & 25.8  & 1.075 & 0.433 & 1507.63 \\
                   & VMC &       & 2.200 & 0.165 & 1251.90 \\
$^{8}$B+$^{208}$Pb & 3CM & 170.3 & 0.661 & 0.389 & 3226.73 \\
                   & VMC &       & 1.358 & 0.908 & 3158.15 \\
\noalign{\smallskip}\hline
\end{tabular}
\end{center}
\end{table}

\begin{figure}
\resizebox{0.46\textwidth}{!}{%
  \includegraphics{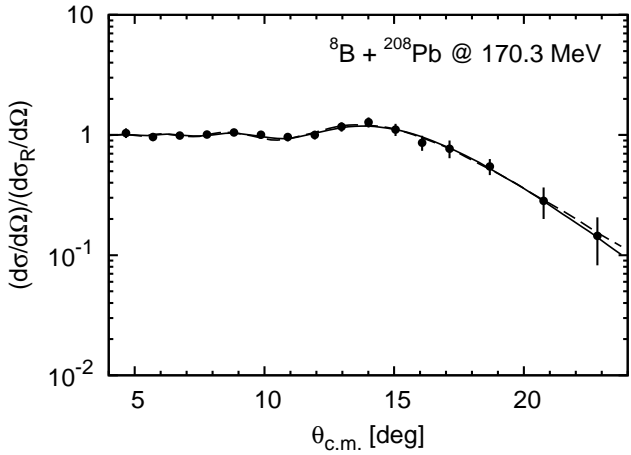}
} \caption{$^{8}$B+$^{208}$Pb elastic scattering cross sections at
$E=170.3$ MeV. Solid line: calculations with 3CM density of
$^{8}$B; dashed line: calculations with VMC density of $^{8}$B.
Experimental data are taken from Ref.~\protect\cite{Yang2013}.}
\label{fig3}
\end{figure}

Our next step is to study $^{8}$B elastic scattering on a lead
target at 170.3 MeV incident energy. Although the experiment was
performed with a natural lead target \cite{Yang2013}, we calculate
the angular distribution as it has been also theoretically
interpreted in Ref.~\cite{Yang2013} assuming a pure $^{208}$Pb
target. Fig.~\ref{fig3} shows a fair agreement of our microscopic
calculations with the experimental data for the cross section.
Both VMC and 3CM densities used in the calculations are able to
reproduce the data that are restricted in a range of small angles.
The values of $N_{R}$ and $N_{I}$ are given in Table~\ref{tab2}.
One can mention that the optical model analysis performed in
Ref.~\cite{Yang2013} on the base of a single-folding model using
the Bruy\`{e}res Jeukenne-Lejeune-Mahaux nucleon-nucleus potential
leads also to a fairly good agreement with experimental data.
Similarly to the case of $^{8}$B+$^{12}$C reaction illustrated in
Fig.~\ref{fig2}, the reasonable agreement of our model with the
data on $^{8}$B+$^{208}$Pb elastic scattering is in favor of the
very weak contribution from other reaction mechanisms, which is
supported by the results from CDCC calculations
\cite{Barioni2011,Yang2013,Paes2012}.

\begin{figure}
\resizebox{0.46\textwidth}{!}{%
  \includegraphics{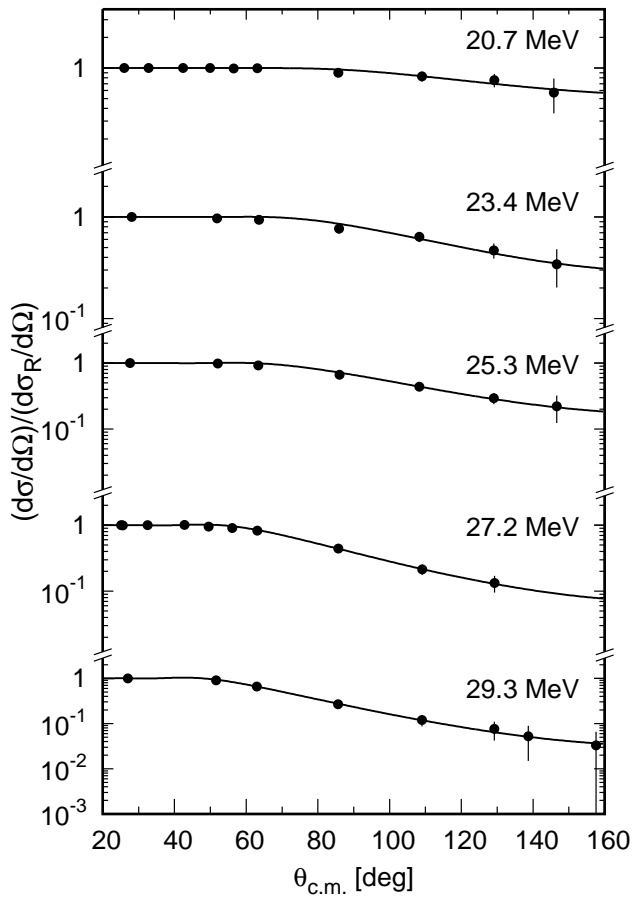}
} \caption{$^{8}$B+$^{58}$Ni elastic scattering cross sections at
$E=20.7, 23.4, 25.3, 27.2$ and 29.3 MeV calculated using the VMC
density of $^{8}$B. Experimental data are taken from
Ref.~\protect\cite{Aguilera2009}.}
\label{fig4}
\end{figure}

\begin{figure}
\resizebox{0.46\textwidth}{!}{%
  \includegraphics{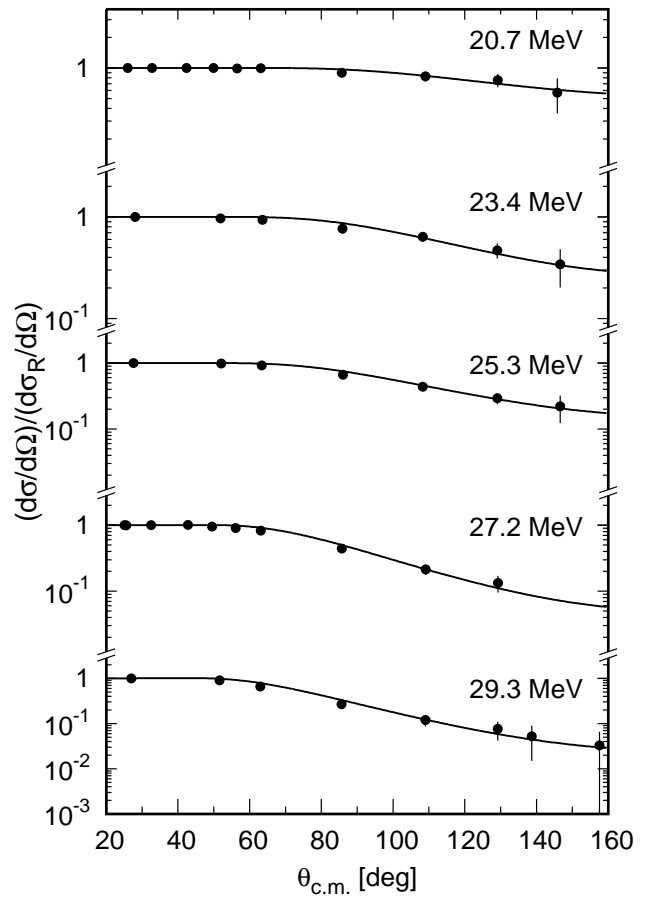}
} \caption{The same as in Fig.~\ref{fig4} but using the 3CM
density of $^{8}$B.}
\label{fig5}
\end{figure}

In what follows, we present in Figs.~\ref{fig4} and \ref{fig5} our
results for $^{8}$B+$^{58}$Ni elastic scattering cross sections at
energies 20.7, 23.4, 25.3, 27.2, and 29.3 MeV using the VMC and
3CM densities, respectively. These results are obtained with
$N_{R}$ and $N_{I}$ which reproduce in a best way the experimental
cross sections at considered five energies. Their values from the
fitting procedure providing minimal $\chi^{2}/N$ are presented in
Table~\ref{tab3}. One can see that the results obtained using both
densities of $^{8}$B are in a good agreement with the data for all
energies considered. The values of the renormalization parameters
$N_{R}$ are smaller than the ones deduced for the $^{8}$B+$^{12}$C
and $^{8}$B+$^{208}$Pb elastic scattering processes (see
Table~\ref{tab2}), in particular when using 3CM density, but the
choice of the parameters is based on the same consistent
$\chi^{2}$ criterion. Here we would like to note that in many
cases of describing real data, the elastic scattering results are
not enough to determine in a unique way the parameters $N_{R}$ and
$N_{I}$. It is well known that the couplings to non-elastic
channels lead to polarization potentials that can considerably
modify the bare potential calculated within the double folding
formalism. Obviously, for more successful description of cross
sections at low energies near Coulomb barrier an inclusion of
polarization contributions due to virtual excitations and decay
channels of the reactions (involving also the fusion data
Ref.~\cite{Camacho2001} to explain correctly the presence of the
breakup threshold anomaly for $^{8}$B+$^{58}$Ni process) is
necessary to obtain unambiguously the OP renormalization
parameters. The good fit obtained for the experimental angular
distributions in Ref.~\cite{Aguilera2009} with real and imaginary
potentials of the Woods-Saxon type and our best fit to the same
data using microscopic OP in this work lead to values of the
predicted total reaction cross section $\sigma_{R}$ very close to
each other, the latter exhibiting a smooth increase with the
energy increase. The good agreement of our results for
$^{8}$B+$^{58}$Ni elastic scattering cross sections with the
experimental data using both VMC and 3CM densities validates their
ability as a reasonable choice to reveal the proton-halo structure
of $^{8}$B nucleus.

\begin{table}
\caption{The renormalization parameters $N_{R}$, $N_{I}$, and the
total reaction cross sections $\sigma_{R}$ (in mb) for results of
the $^{8}$B+$^{58}$Ni elastic scattering at five incident energies
$E$ (in MeV) shown in Figs.~\ref{fig4} and \ref{fig5} using the
VMC and 3CM model densities of $^{8}$B, respectively.}
\label{tab3}
\begin{center}
\begin{tabular}{llllll}
\hline\noalign{\smallskip}
Model & $E$ & $N_R$ & $N_I$ & $\sigma_R$\\
\noalign{\smallskip}\hline\noalign{\smallskip}
VMC & 20.7  & 0.863 & 2.792 & 208.69  \\
3CM &       & 0.427 & 0.393 & 201.99  \\
VMC & 23.4  & 0.359 & 1.500 & 376.08  \\
3CM &       & 0.319 & 0.266 & 370.04  \\
VMC & 25.3  & 0.317 & 1.030 & 494.86  \\
3CM &       & 0.235 & 0.212 & 480.38  \\
VMC & 27.2  & 0.329 & 1.750 & 794.03  \\
3CM &       & 0.293 & 0.252 & 705.38  \\
VMC & 29.3  & 0.525 & 1.830 & 987.29  \\
3CM &       & 0.221 & 0.248 & 847.11  \\
\noalign{\smallskip}\hline
\end{tabular}
\end{center}
\end{table}

\begin{figure}
\resizebox{0.46\textwidth}{!}{%
  \includegraphics{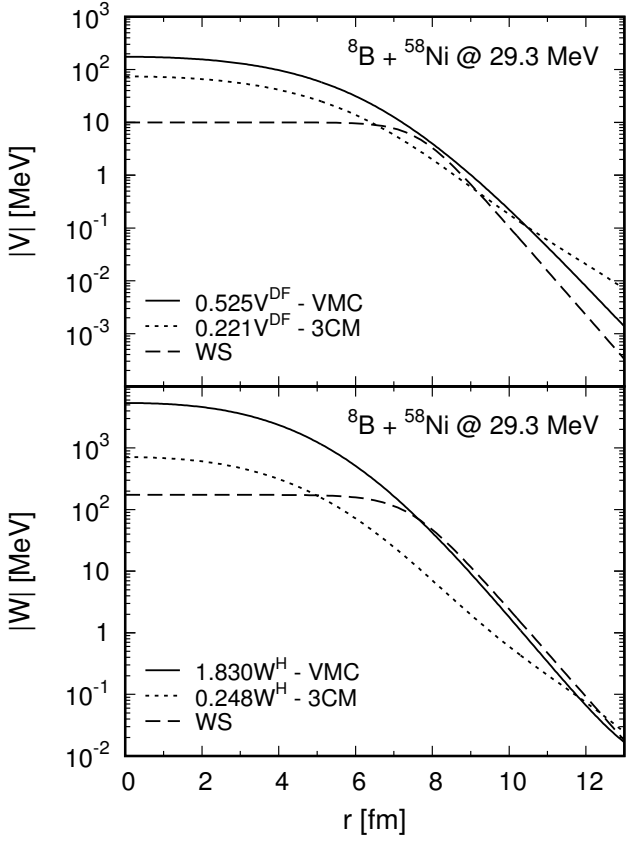}
} \caption{The absolute values of the real $N_{R}V^{DF}$ and
imaginary $N_{I}W^{H}$ parts of the calculated optical potentials
for the $^{8}$B+$^{58}$Ni elastic scattering at $E=29.3$ MeV
obtained using the VMC and 3CM densities of $^{8}$B in comparison
with those of the WS potential from Ref.~\cite{Aguilera2009}. The
values of $N_{R}$ and $N_{I}$ are from Table~\ref{tab3}.}
\label{fig6}
\end{figure}

\begin{figure}[t]
\resizebox{0.46\textwidth}{!}{%
  \includegraphics{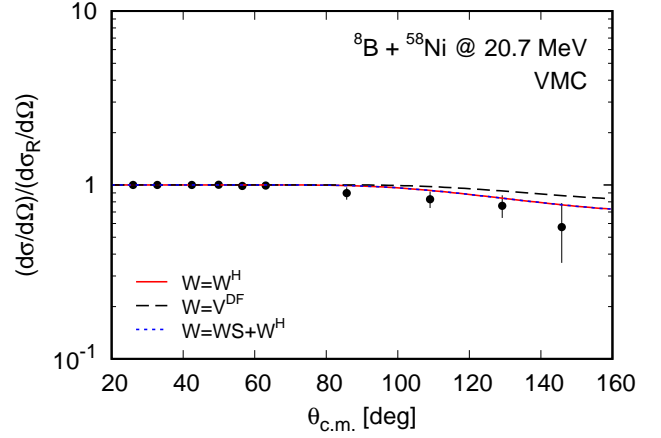}
} \caption{$^{8}$B+$^{58}$Ni elastic scattering cross sections at
$E=20.7$ MeV calculated using the VMC density of $^{8}$B. Red
solid line: $W=W^{H}$; black dashed line: $W=V^{DF}$; blue dotted
line: $W=WS+W^{H}$ (see also the text). Experimental data are
taken from Ref.~\protect\cite{Aguilera2009}.}
\label{fig7}
\end{figure}

Further, we give in Fig.~\ref{fig6}, as an example (for $E=29.3$
MeV), the comparison of the obtained real and imaginary parts of
the OPs for both 3CM and VMC densities with the corresponding
parts of the fitted Woods-Saxon potential \cite{Aguilera2009}. The
values of our parameters $N_{R}$ and $N_{I}$ are those from
Table~\ref{tab3}. Here we mention that at such energies the
surface part of the ImOP plays a decisive role on the behavior of
the elastic cross sections. One can see from Fig.~\ref{fig6} that
the use of the VMC density leads to a very good agreement of the
imaginary part of our OP with the imaginary part of the fitted WS
OP in the surface region. Also, the slope of the real part of OP
obtained with the VMC density in this region ($8<r<10$ fm) is
similar to that of the real part of WS OP. There exist some
differences in the surface region for the real and imaginary parts
of the OP obtained with the 3CM density and the corresponding
parts of the WS OP.

Along this line, we present in Fig.~\ref{fig7} the results of
another calculation, namely of the elastic $^{8}$B+$^{58}$Ni cross
section (e.g., at energy $E=20.7$ MeV). The three curves
correspond to the real part of the OP $V^{DF}$ from the double
folding procedure, while the ImOP is calculated in three ways: i)
from HEA ($W=W^{H}$); ii) equal to the real part of the OP
($W=V^{DF}$), and iii) the central part of the ImOP is taken to be
in a form of a WS potential up to $r=7$ fm, while the surface
component of the ImOP (at $r>7$ fm) is taken to be equal to that
from HEA ($W=W^{H}$). The parameters of WS potential are given in
Table~I of Ref.~\cite{Aguilera2009} and its depth at $r=0$ fm is
around two times smaller than $V^{DF}(r=0)$.  It can be seen that
all three curves are close to each other that shows the importance
of the surface part of the OP which is similar in all three cases.
In this way, in the further considerations we do not pay attention
to the non-realistic values of the central part of the HEA ImOP
($W=W^{H}$) and are concentrated on the surface component of this
potential.

In the context of the obtained results for the $^{8}$B+$^{58}$Ni
elastic scattering at near-Coulomb barrier energies, we would like
also to mention the contributions of other mechanisms that play a
crucial role. Lubian {\it et al.} showed in Ref.~\cite{Lubian2009}
that taking into account the coupling between elastic and breakup
channels, the CDCC calculations reproduce very well the data of
Ref.~\cite{Aguilera2009}. Although the inclusion of
continuum-continuum couplings were essential to have a good
agreement with the data, it was noted that nuclear excitations of
the target have a weak influence on the elastic angular
distributions. In addition, the analysis carried out in
Ref.~\cite{Camacho2001} led to a conclusion that the breakup
threshold anomaly is present for the $^{8}$B+$^{58}$Ni system at
energies close to the Coulomb barrier ($V_{B}=20.8$ MeV). So, all
these findings support the important role played by the
Coulomb-nuclear interference at large distances for a halo nucleus
as $^{8}$B.


\section{Breakup reactions of $^{8}$B}
\label{s:theorybu}
\subsection{The $^{7}$Be+$p$ model of $^{8}$B}
\label{s:model}

In this section we consider the characteristics of breakup
processes of the $^{8}$B nucleus on the example of the stripping
reaction cross sections and the momentum distributions of the
fragments. We use a simple model in which $^{8}$B consists of a
core of $^{7}$Be and a halo of a single proton (see, e.g.,
Refs.~\cite{Smedberg99,Schwab95}). In this model the density of
$^{7}$Be has to be known. We use that one obtained from the
calculations performed by means of the 3CM density of $^{8}$B
\cite{Varga95}. Later in this subsection we give the expressions
how to obtain the corresponding $S$-matrices needed to calculate
the breakup cross sections through the imaginary part of
cluster-target OPs within the HEA. The wave function of the
relative motion of the proton and $^{7}$Be clusters in $^{8}$B is
obtained by solving the Schr\"{o}dinger equation with the
Woods-Saxon potential for a particle with a reduced mass of two
clusters. The parameters of the WS potentials are obtained by a
fitting procedure, namely, to reach the proton separation energy
$S_{p}=137$ KeV. However, this procedure could provide several
sets of potential parameters that satisfy the above condition.
They are close to each other leading, at the same time, to
different valence proton rms radii. Therefore, in order to
understand better the observed widths of the longitudinal momentum
distributions of $^{7}$Be fragments formed in the breakup of
$^{8}$B and measured at different targets and energies, we
consider three cases. The values of WS potential parameters and
corresponding rms radii of the cluster formation for $1p$ state in
which the valence proton in $^{8}$B is mainly bound (see
Refs.~\cite{Hencken96,Esbensen96}) are listed in Table~\ref{tab4}.

\begin{table}
\caption{The parameters $V_{0}$ (in MeV), $R$ (in fm), $a$ (in fm)
of the Woods-Saxon potentials and the rms radii of the cluster
wave function (in fm) obtained by using of the 3CM density of
$^{7}$Be for three cases (see the text).}
\label{tab4}
\begin{center}
\begin{tabular}{lllll}
\hline\noalign{\smallskip}
$V_{0}$ & $R$ & $a$ & rms radii\\
\noalign{\smallskip}\hline\noalign{\smallskip}
38.22 & 2.70 & 0.55 & 4.51 \\
38.70 & 2.50 & 0.20 & 5.08 \\
38.77 & 2.48 & 0.50 & 6.24 \\
\noalign{\smallskip}\hline
\end{tabular}
\end{center}
\end{table}

The wave function of the $p$-state ($l=1$) of the relative motion
of two clusters has the form
\begin{equation}
\phi_{1m}({\bf s})=R_{1}(s)Y_{1m}(\theta,\varphi),
\label{eq:6}
\end{equation}
where $R_{1}(s)$ is the radial wave function and
$Y_{1m}(\theta,\varphi)$ are the spherical functions for $l=1$.
The corresponding probability density of both clusters to be at a
mutual distance $s$ is written as
\begin{equation}
\rho_{0}({\bf s})=\frac{1}{4\pi}|\phi_{1m}({\bf s})|^{2}.
\label{eq:8}
\end{equation}

For calculations of breakup cross sections and momentum
distributions of fragments in the $^{7}$Be+$p$ breakup model
within the eikonal formalism (see, e.g. Ref.~\cite{Hencken96}),
one needs the expressions of the $S$-matrix (as a function of the
impact parameter $b$):
\begin{equation}
S(b)=\exp \left [ -\frac{i}{\hbar v}\int_{-\infty}^{\infty} U^{(b)}(\sqrt{b^{2}+z^{2}})dz \right ],
\label{eq:10}
\end{equation}
where
\begin{equation}
U^{(b)}=V+ i W
\label{eq:11}
\end{equation}
is the OP of the breakup of $^{8}$B in its collision with nuclear
targets within the $^{7}$Be+$p$ cluster model. Correspondingly
(for negative $W$), one can write
\begin{equation}
|S(b)|=\exp \left [ -\frac{1}{\hbar v}\int_{-\infty}^\infty |W| dz \right ].
\label{eq:12}
\end{equation}
$|S(b)|^2$ gives the probability that after the collision with a
target nucleus ($z\rightarrow \infty$), the cluster $c$ or the
proton with impact parameter $b$ remains in the elastic channel
($i=c,p$):
\begin{equation}
|S_{i}(b)|^{2}=\exp{\left[-\frac{2}{\hbar
v}\int_{-\infty}^{\infty} dz\,\left |W_i
(\sqrt{b^{2}+z^{2}})\right |\right ]},
\label{eq:13}
\end{equation}
where $W_{c}$ and $W_{p}$ are the imaginary parts of the OP
(\ref{eq:11}) of $^{7}$Be+$A$ and $p+A$ scattering, respectively.
They are calculated microscopically using the procedure given in
Subsec.~\ref{s:op}. The probability a cluster to be removed from
the elastic channel is $(1-|S|^{2})$. The probability of the case
when both clusters ($c$ and $p$) leave the elastic channel is
$(1-|S_{p}|^{2})(1-|S_{c}|^{2})$.

The longitudinal momentum distribution of $^{7}$Be fragments
produced in the breakup of $^{8}$B in the case of stripping
reaction (when the proton leaves the elastic channel) is
\begin{eqnarray}
\left(\frac{d\sigma}{dk_{L}}\right)_{str}
&=&\frac{1}{3}\int_{0}^{\infty}b_{p}db_{p}
\left [ 1-|S_{p}(b_{p})|^{2}\right ] \nonumber \\
& \times & \int \rho d\rho d\varphi_{\rho} |S_{c}(b_{c})|^{2}
\sum_{m=0,\pm 1} F_{m}(\rho),
\label{eq:14}
\end{eqnarray}
where
\begin{equation}
F_{m}(\rho)=\left |\int_{-\infty}^{\infty} dz \exp
(-ik_{L}z)Y_{1m}(\theta,\varphi)R_{1}(r)\right |^{2}.
\label{eq:15}
\end{equation}
After substituting the spherical functions
$Y_{1m}(\theta,\varphi)$ in Eq.~(\ref{eq:15}) one obtains the
following expressions
\begin{equation}
F_{0}=\frac{3}{\pi}\left [\int_{0}^{\infty} dz \sin (k_{L}z)
\frac{z}{\sqrt{\rho^{2}+z^{2}}} R_{1}(\sqrt{\rho^{2}+z^{2}})
\right ]^{2},
\label{eq:16}
\end{equation}
\begin{eqnarray}
F_{1}&=&F_{+1}=F_{-1} \nonumber \\
&=& \frac{3}{2\pi}\left [\int_{0}^{\infty} dz \cos (k_{L}z)
\frac{\rho}{\sqrt{\rho^{2}+z^{2}}} R_{1}(\sqrt{\rho^{2}+z^{2}})
\right ]^{2}.
\label{eq:17}
\end{eqnarray}
Equation (\ref{eq:14}) is obtained when the incident nucleus has
spin equal to zero and for the $p$-state of the relative motion of
two clusters in the nucleus with ${\bf s}={\bf r}_{c}- {\bf
r}_{p}$, ${\bf \rho}={\bf b}_{c}-{\bf b}_{p}$, ${\bf s}={\bf \rho}
+ {\bf z}$ and
\begin{equation}
b_{c}=\sqrt{s^{2}\sin^{2}\theta+b_{p}^{2}+2sb_{p}\sin\theta\cos(\varphi-
\varphi_{p})}
\label{eq:18}
\end{equation}
coming from ${\bf b}_{c}={\bf b}_{p}+{\bf b}$, where
$b=s\sin\theta$ is the projection of ${\bf s}$ on the plane normal
to the $z$-axis along the straight-line trajectory of the incident
nucleus.

\subsection{Results of calculations of breakup reactions}
\label{s:calculationsbu}

In this subsection we perform calculations of the breakup cross
sections of $^{8}$B on the target nuclei $^{9}$Be, $^{12}$C, and
$^{197}$Au and compare our results with the available experimental
data \cite{Kelley96,Jin2015}. The densities of these nuclei needed
to compute the OPs are taken from Ref.~\cite{Pieper2015} for
$^{9}$Be, Ref.~\cite{Burov77} for $^{12}$C, and
Ref.~\cite{Patterson2003} for $^{197}$Au, respectively.

\begin{figure}
\resizebox{0.46\textwidth}{!}{%
  \includegraphics{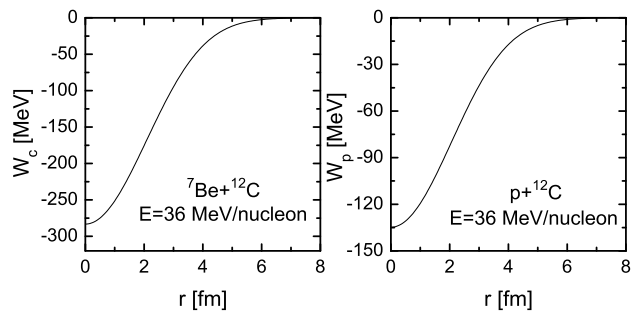}
} \caption{Imaginary parts $W_{c}$ and $W_{p}$ of cluster-target
potentials for calculation of $^{8}$B breakup on $^{12}$C target
at $E$=36 MeV/nucleon.}
\label{fig9}
\end{figure}

In our work, to calculate the corresponding $S$-matrices
[Eqs.~(\ref{eq:12}) and (\ref{eq:13})] one needs only the
imaginary part of cluster-target potentials to be known. As an
example, we illustrate in Fig.~\ref{fig9} the imaginary parts of
both $^{7}$Be+$^{12}$C and $p$+$^{12}$C OPs calculated
microscopically in HEA by using Eq.~(\ref{eq:5}) within the
cluster model. The results given in Fig.~\ref{fig9} are obtained
at the same energy of 36 MeV/nucleon as the longitudinal momentum
distributions of $^{7}$Be fragments in the breakup of $^{8}$B on a
carbon target have been measured \cite{Jin2015}.
It is worth noting that although there are no $^{7}$Be+$^{12}$C
and $^{8}$B+$^{12}$C elastic scattering data available at this
specific energy, the diffraction nuclear model developed in
Ref.~\cite{Kovalchuk2009} reproduces the elastic scattering of
projectile exotic nuclei $^{7}$Be and $^{8}$B by $^{12}$C at
similar energy (40 MeV/nucleon) \cite{Pecina95}. In
\cite{Kovalchuk2009} the interaction of $^{7}$Be as a formation of
two clusters ($\alpha$+$^{3}$He) is accounted for explicitly. If
measurements of the cluster-target ($^{7}$Be+$^{12}$C and
$p$+$^{12}$C) elastic scattering at energy 36 MeV/nucleon exist,
one should provide additional information on the properties not
only of the imaginary parts of the corresponding core-target and
proton-target potentials (shown in Fig.~\ref{fig9}), but also of
their real parts (although they are not needed to calculate the
breakup processes in the present work).

\begin{figure}[b]
\resizebox{0.47\textwidth}{!}{%
  \includegraphics{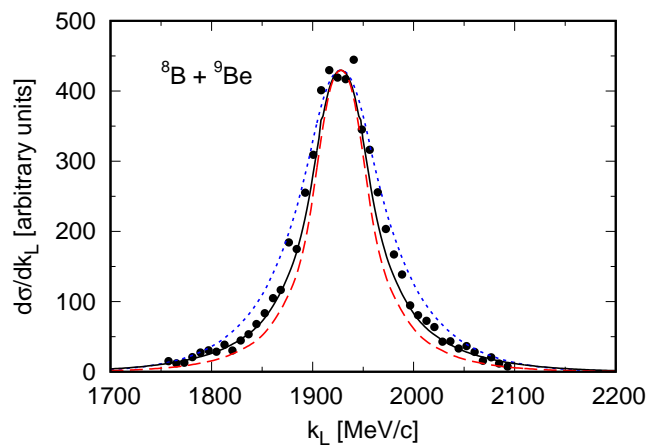}
} \caption{Cross section of stripping reaction in $^{8}$B+$^{9}$Be
scattering at $E$=41 MeV/nucleon. Experimental data are taken from
Ref.~\cite{Kelley96}.}
\label{fig10}
\end{figure}

\begin{figure}
\resizebox{0.46\textwidth}{!}{%
  \includegraphics{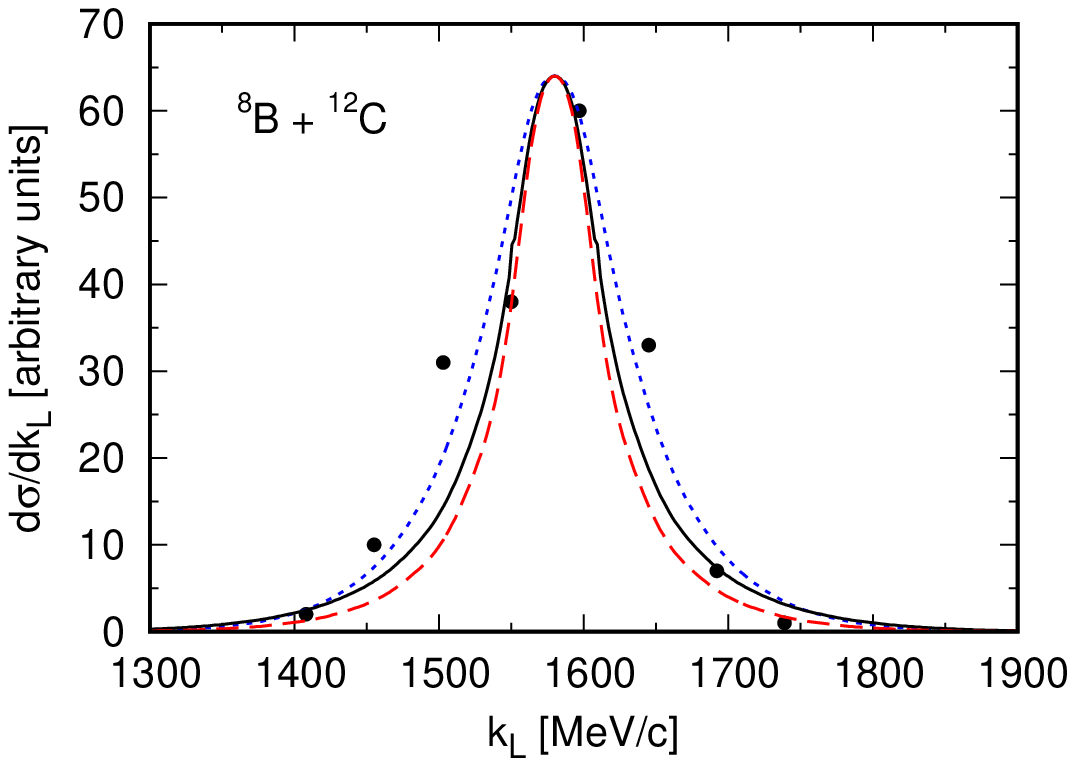}
} \caption{Cross section of stripping reaction in $^{8}$B+$^{12}$C
scattering at $E$=36 MeV/nucleon. Experimental data are taken from
Ref.~\cite{Jin2015}.}
\label{fig11}
\end{figure}

\begin{figure}
\resizebox{0.47\textwidth}{!}{%
  \includegraphics{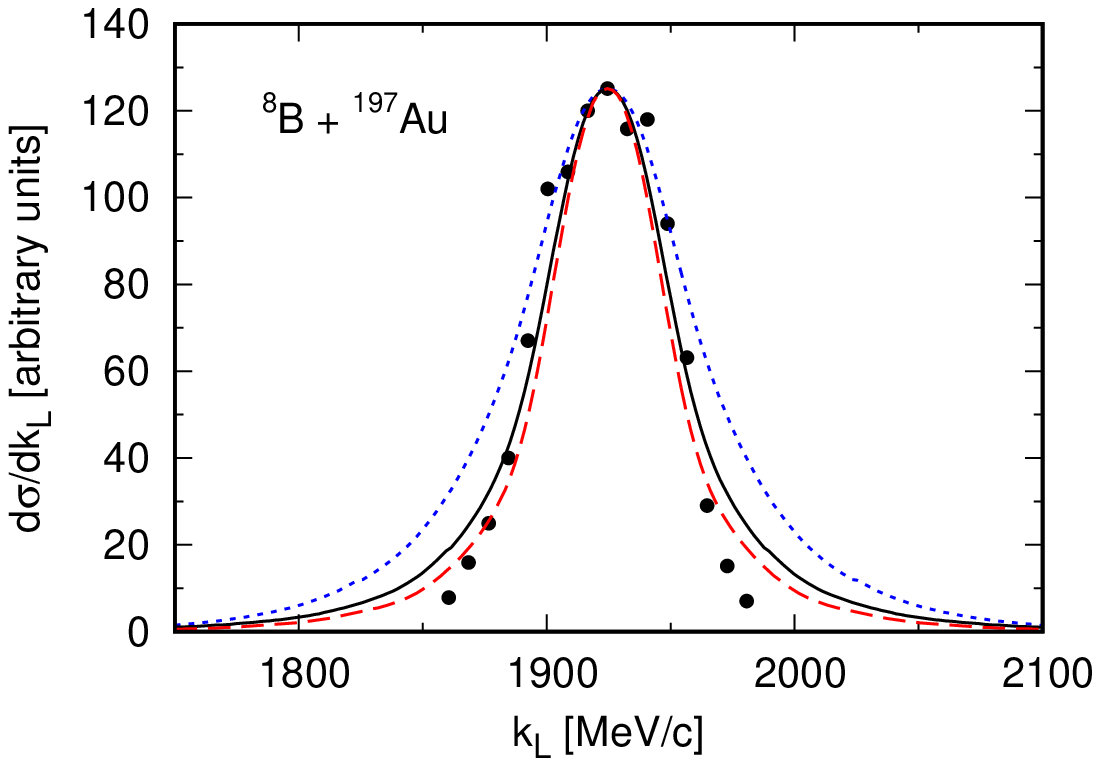}
} \caption{Cross section of stripping reaction in
$^{8}$B+$^{197}$Au scattering at $E$=41 MeV/nucleon. Experimental
data are taken from Ref.~\cite{Kelley96}.}
\label{fig12}
\end{figure}

The stripping cross sections (when a proton leaves the elastic
channel) for reactions $^{8}$B+$^{9}$Be, $^{8}$B+$^{12}$C, and
$^{8}$B+$^{197}$Au are calculated from Eq.~(\ref{eq:14}). The
obtained results are illustrated in Figs.~\ref{fig10},
\ref{fig11}, and \ref{fig12}, respectively. The blue dotted, black
solid, and red dashed curves in the figures correspond to the
three sets of WS parameters given in Table~\ref{tab4}. Here we
note that due to the arbitrary units of the measured cross
sections of the considered processes it was not necessary to
renormalize the depths of our OPs of the fragments-target nuclei
interactions. It is worth to note the relevance between the rms
radii of the wave function of the $^{7}$Be-$p$ relative motion and
the obtained FWHMs for the considered three cases. The latter are
presented in Table~\ref{tab5} together with their experimental
values. Due to the uncertainty principle the widths become smaller
with the increase of the distance between two clusters. We note
the good agreement with the experimental data from light and heavy
breakup targets. It can be seen from Figs.~\ref{fig10} and
\ref{fig11} that the best agreement with the experimental data for
the parallel momentum distributions of $^{7}$Be fragments in a
breakup reaction of $^{8}$B on a $^{9}$Be target at 41 MeV/nucleon
and on a $^{12}$C target at 36 MeV/nucleon is achieved when the
relative $^{7}$Be-proton distance is 5.08 fm or 4.51 fm,
respectively, while in the case of $^{8}$B breakup on a $^{197}$Au
target at 41 MeV/nucleon shown in Fig.~\ref{fig12} a larger
distance (6.24 fm) is needed to get better coincidence with the
data. This observation is proved by the performed $\chi^{2}$
analysis for the deviation of the theoretical results from the
data. The necessity to take into account the Coulomb distortion of
straight-line trajectories for the heavier nucleus when
calculating phases in Eqs.~(\ref{eq:12}) and (\ref{eq:13}) is a
reason for getting larger distance in the case of $^{8}$B breakup
on a $^{197}$Au target. The obtained FWHMs that correspond to
these rms radii listed in Table~\ref{tab4} are close to the
experimentally measured widths. In addition, our FWHM values are
within the range found in other theoretical analyses, for
instance, 103 MeV/c and 107 MeV/c when describing the stripping
mechanism of $^{8}$B breakup on $^{12}$C target (denoted KDe and
KDp in Ref.~\cite{Jin2015}), 55 MeV/c and 61 MeV/c FWHM from the
analysis of $^{8}$B breakup on $^{197}$Au target \cite{Kelley96},
and width of about 100 MeV/c of the momentum distribution of
$^{7}$Be fragments from the breakup of $^{8}$B on $^{9}$Be target
\cite{Kelley96} when a stripping model \cite{Esbensen96} was
employed.

Here we would like to discuss shortly the applicability of HEA to
energies considered in the breakup processes. As known generally,
the HEA is applied to energies larger than 100 MeV/nucleon.
However, in the last years the HEA was modified and applied also
to lower energies (see, e.g.,
Refs.~\cite{Lukyanov2004,Shukla2003,Vries80,Vitt87,Charagi92,Charagi97,Brink81,Vismes2000,Lukyanov2007,Lukyanov2009}).
The prescription to calculate the profile function in this case
consists in a replacement of the straight-line trajectory impact
parameter ($b$) by the distance of the closest approach ($b_{c}$)
in the Coulomb field, or by the respective distance ($r_{cn}$) in
the presence of nuclear field (ReOP). Recently, the eikonal
description of the breakup of exotic nuclei has been developed at
low energies by including the Coulomb deflection of the projectile
off the target \cite{Fukui2014,Capel2016}. In principle, this
correction is based on the same replacement of the impact
parameter $b$ by $b_{c}$ mentioned above and the results obtained
in Ref.~\cite{Capel2016} for the breakup of $^{15}$C on Pb at 20
MeV/nucleon confirm the ability of the eikonal approximation to be
reliably extended to low energies.

Concerning the use of HEA in the present work, we should note that
the fragment-folding potentials $W_{i}(r)$ in Eq.~(\ref{eq:13})
are rather deep since they are proportional to the number of
nucleons in the target- and fragment-nucleus with radii $R_{t}$
and $R_{i}$, respectively. Therefore, a very strong absorption
takes place at $r<R=R_{t}+R_{i}$, and so only the surface region
of $W$ plays decisive role in the process. Thus, the condition for
applicability of the eikonal approach can be estimated as
$EA_{i}\gg W(r\cong R_t + R_i)$ ($E$ being energy per nucleon)
that is fulfilled in our calculations.

In the end, we note that from the comparison of the results in the
present work with those obtained in our previous works
\cite{Lukyanov2013,Lukyanov2015} for the cases of $^{11}$Li and
$^{11}$Be breakup, it can be concluded that the halo cluster ($2n$
or $n$) in these nuclei can be found outside the core with larger
probability than the valence proton in $^{8}$B. This observation
together with the variation of the width with target in breakup
reactions of $^{8}$B at almost equal energies show the specific
features of the momentum distributions of corelike fragments in
breakup of neutron and proton halo nuclei.
\begin{table}
\caption{The values of theoretical and experimental FWHM for
stripping mechanism of the $^{8}$B breakup on $^{9}$Be, $^{12}$C,
and $^{197}$Au targets (in MeV/c) at incident energy $E$=41, 36,
and 41 MeV/nucleon, respectively. The order of the results for
FWHM obtained in this work corresponds to the blue dotted, black
solid, and red dashed lines given in Figs.~\ref{fig10},
\ref{fig11}, and \ref{fig12} and rms radii listed in
Table~\ref{tab4}.}
\label{tab5}
\begin{center}
\begin{tabular}{llll}
\hline\noalign{\smallskip}
Process & Present work & Exp. \\
\noalign{\smallskip}\hline\noalign{\smallskip}
$^{8}$B+$^{9}$Be   &  95.55 & $81\pm 4$   \\
                   &  72.07 &             \\
                   &  63.21 &             \\
$^{8}$B+$^{12}$C   & 108.71 & $124\pm 17$ \\
                   &  81.91 &             \\
                   &  70.02 &             \\
$^{8}$B+$^{197}$Au &  79.64 & $62\pm 3$   \\
                   &  61.14 &             \\
                   &  54.86 &             \\
\noalign{\smallskip}\hline
\end{tabular}
\end{center}
\end{table}

\section{Summary and conclusions}
\label{s:conclusions}

In the present work we performed microscopic calculations of the
optical potentials and cross sections of elastic scattering
$^{8}$B+$^{12}$C at 25.8 MeV, $^{8}$B+$^{58}$Ni at 20.7, 23.4,
25.3, 27.2, and 29.3 MeV, and $^{8}$B+$^{208}$Pb at 170.3 MeV, in
comparison with the available experimental data. The direct and
exchange isoscalar and isovector parts of the real OP ($V^{DF}$)
were calculated microscopically using the double-folding procedure
and density dependent M3Y (CDM3Y6-type) effective interaction
based on the Paris $NN$ potential. The imaginary part of the OP
($W^{H}$) was calculated as a folding integral that corresponds to
the one in a phase of HEA and also in a form of the
microscopically calculated ReOP ($W=V^{DF}$), where
antisymmetrization is taken into account. Two model microscopic
densities of protons and neutrons in $^{8}$B were used in the
calculations: the density calculated within the VMC model
\cite{Carlson2015,Pieper2015} and from the three-cluster model
\cite{Varga95}. The nucleon density distributions of $^{12}$C
\cite{Burov77}, $^{58}$Ni \cite{Khoa2000}, and $^{208}$Pb
\cite{Patterson2003} were taken as defolded charge densities
obtained from the best fit to the experimental form factors from
electron elastic scattering on these nuclei. The elastic
scattering differential cross sections and total reaction cross
sections were calculated using the program DWUCK4 \cite{DWUCK}. In
this way, in contrast to the phenomenological and semi-microscopic
models we deal with a fully microscopic approach as a physical
ground to account for the single-particle structure of the
colliding nuclei.

It turned out that the values of the coefficients $N_{R}$ and
$N_{I}$ in Eq.~(\ref{eq:1}) that renormalize the ReOP and ImOP
depend on the density of $^{8}$B used in the calculations. The use
of the VMC and 3CM densities leads to good agreement with the
experimental cross sections. An unambiguous determination of
$N_{R}$ and $N_{I}$ of our microscopic OP, especially if one
analyzes the reaction $^{8}$B+$^{58}$Ni, can be achieved when a
simultaneous fit to all data corresponding to different reaction
mechanisms is performed \cite{Camacho2001}. The analysis of the
behavior of VMC and 3CM densities and the corresponding OPs (see
Fig.~\ref{fig6}) in comparison with the fitted WS OP from
Ref.~\cite{Aguilera2009}, as well as our results shown in
Fig.~\ref{fig7} give additional information on the decisive role
of the nuclear surface on the mechanism of the considered
scattering processes.

We have tested our microscopic model studying the role of the
breakup mechanism to analyze properly the whole picture of $^{8}$B
scattering. For this purpose, we use another folding approach to
consider the $^{8}$B breakup by means of the simple $^{7}$Be+$p$
cluster model for the structure of $^{8}$B. We calculate in HEA
the ImOP of the interaction of $^{7}$Be with the target, as well
as the $p$+target interaction. Using them the corresponding
$S$-matrices for the core and proton within the eikonal formalism
are obtained. The latter are used to get results for the
longitudinal momentum distributions of $^{7}$Be fragments produced
in the breakup of $^{8}$B on different targets. This includes the
breakup reactions of $^{8}$B on $^{9}$Be and $^{197}$Au at $E=41$
MeV/nucleon and $^{8}$B on $^{12}$C at $E=36$ MeV/nucleon, for
which a good agreement of our calculations for the stripping
reaction cross sections with the available experimental data were
obtained. The theoretical widths are close to the empirical ones.

In general, we can conclude that our microscopic approach can be
applied to reaction studies with exotic nuclei such as $^{8}$B.
The consistency of our results with the measured elastic cross
sections and narrow longitudinal momentum distributions may
provide supplemental information on the internal spatial structure
of the  $^{8}$B nucleus supporting its proton-halo nature.

\begin{acknowledgement}
The authors are grateful to S.C. Pieper for providing with the
density distributions of $^{8}$B nucleus calculated within the VMC
method and to S.L. Jin for the experimental longitudinal momentum
distributions of $^{7}$Be fragments from the breakup of $^{8}$B on
a carbon target. The work is partly supported by the Project from
the Agreement for co-operation between the INRNE-BAS (Sofia) and
JINR (Dubna). Four of the authors (D.N.K., A.N.A., M.K.G. and
K.S.) are grateful for the support of the Bulgarian Science Fund
under Contract No.~DFNI--T02/19 and one of them (D.N.K.) under
Contract No.~DFNI--E02/6.
\end{acknowledgement}

\end{document}